\documentclass{article}
\usepackage{spconf,amsmath,graphicx}
\usepackage{subfigure}
\usepackage{multirow}

\title{Stream Attention for far-field multi-microphone ASR}
\name{Xiaofei Wang$^{\dagger, \star}$ Yonghong Yan $^{\star}$ and Hynek Hermansky$^\dagger$ \thanks{This work is supported by the National Natural Science Foundation of China (No. 61601453) and the China Scholarship Council (No. 201604910007)}}
\address{$^\dagger$Center for Language and Speech Processing, Johns Hopkins University \\3400 North Charles Street, Baltimore, MD 21218, USA\\
	$^\star$Institute of Acoustics, Chinese Academy of Sciences \\No.21 North 4th Ring West Road, Beijing 100190, China}
\begin{document}
%\ninept
%
\maketitle
\begin{abstract}
A stream attention framework has been applied to the posterior probabilities of the deep neural network (DNN) to improve the far-field automatic speech recognition (ASR) performance in the multi-microphone configuration. The stream attention scheme has been realized through an attention vector, which is derived by predicting the ASR performance from the phoneme posterior distribution of individual microphone stream, focusing the recognizer's attention to more reliable microphones. Investigation on the various ASR performance measures has been carried out using the real recorded dataset. Experiments results show that the proposed framework has yielded substantial improvements in word error rate (WER).
\end{abstract}
\section{Introduction}
\label{sec:intro}
In far-field ASR scenario, it is feasible to use many parallel recognition streams. A situation needs to be solved where a number of microphones, which form acoustic streams, are distributed in space to acquire speech to be recognized. Depending on the room situation and microphone status, some streams (microphones closer to the speaker, less noise and reverberation) may deliver better recognition results than the others. Automatically selecting the best microphone for ASR, and further achieving a potential better ASR performance through combining the microphones is desirable. Conventional solutions such as selecting the acoustic stream with the highest energy are vulnerable to strong noises. 

There are several ways to enhance the ASR performance utilizing the multi-microphone configuration. One possible strategy is to align the time delay between the microphones and use spatial information to carry out beamforming \cite{zhang2008maximum}\cite{markovich2015optimal}. However, in the distributed setup, time delays are difficult to estimate. Further, as a front-end processing module, the objective functions are not optimal for ASR \cite{meng2017deep}. Another way of approaching this problem is to find the highest likelihood combination of best paths through multiple recognition lattices, formed from all individual streams \cite{fiscus1997post}\cite{xu2010improved}. 
This requires carrying out full searches in each microphone stream, which is typically done over the whole length of each utterance. And the computing complexity of the multiple decoding operations is the bottleneck.

Most ASR systems require feature vectors, which represent information about underlying speech sound at regular time intervals. Such feature vectors can be derived from posterior probabilities of the sounds, estimated by DNN classifiers. DNN posteriors are able to tolerate the misalignment between the classifier inputs and corresponding labels \cite{waibel1989phoneme}. We propose to construct at every time instant the best feature vector from a combination of the most reliable sound posteriors from different available streams, which is a stream attention scheme. In this way, only one decoding operation for ASR is needed, which is more effective than multiple operations.

Attention scheme can be achieved by generating an attention vector for multiple inputs \cite{bahdanau2014neural}\cite{kim2016recurrent}, among which the attention vector plays an important role in addressing the crucial part of the inputs based on specific attention criterion. Given the feature vectors (DNN posteriors), the key problem of stream attention that to deal with is to find an appropriate measure of the goodness of feature vectors in the individual streams. This goodness measure could then be used in deriving proper attention vector for the construction of the best feature vector. 

In this study, we propose a stream attention framework to deal with the far-field multi-microphone ASR problem, in which the sounds from microphones are not forced aligned. For better understanding the framework,
we investigate several measures that built the relationship between the goodness of DNN posterior vectors and the ASR performance \cite{okawa1998multi}\cite{misra2002entropy}\cite{hermansky2013mean}\cite{meyer2016performance}\cite{mallidi2015uncertainty}\cite{mallidi2015autoencoder}, 
and test the framework using various attention vectors on a real recorded dataset. Specifically, attention vectors are estimated based on the discriminative judgment among the microphone streams.
%Further, We investigate the robustness of the framework, given a particular acoustic perturbation that some microphone streams almost complete break down of the recognition process.

The remainder of this paper is organized as follows: Section 2 describes the proposed stream attention framework of the multi-microphone system. In section 3, different ASR performance measures are compared in far-field multi-microphone ASR experiments using real recordings. Section 4 concludes the paper.

\section{Proposed Framework}
\label{sec:system}
\begin{figure}[htb]
  \centering
  \centerline{\includegraphics[width=8.5cm]{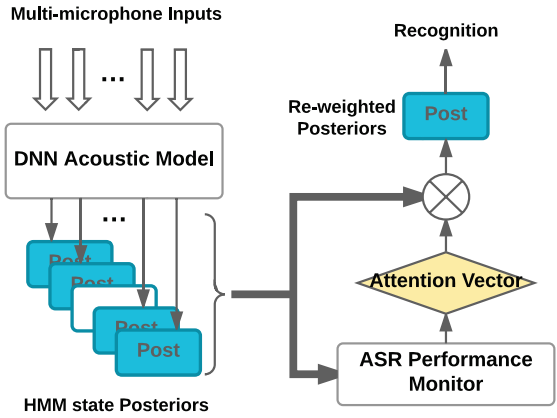}}
\caption{Stream attention framework for ASR using the posterior probabilities form DNN classifier.}
\label{fig:res}
\end{figure}

In this section, we describe the stream attention framework applied on the posterior probabilities of Hidden Markov Model (HMM) state to force the recognizer automatically focusing on the reliable microphones in the multi-microphone configuration. A brief diagram in Fig.\ref{fig:res} demonstrates the attention scheme and attention vector estimation using the multiple posteriors, with each corresponding to the Softmax output of a typical DNN-HMM classifier.

\subsection{Formulation of the Stream Attention Scheme}

As suggested in Fig.\ref{fig:res}, let ${\bf P}_t=[P_t^1,P_t^2,...,P_t^M]^T$ denote the posterior probability sequences of HMM states $O$ at time $t$, where ${}^T$ is the transpose operation and $P_t^i=p(O|{\bf X}_t^i),i=1,...,M$ is the $i$th posterior probability sequence given the feature sequence ${\bf X}_t^i$ extracted from the signal of microphone $i$. $M$ is the total stream number, which is equal to the number of microphones (or arrays). Specifically, ${\bf X}_t^i=[X_{t-\tau}^i,...,X_t^i,...,X_{t+\tau}^i]^T$ is context based, including $2\tau+1$ adjacent frames centered at time $t$.

Assuming that we have the stream attention vector ${\bf w}_t=[w_t^1,w_t^2,...,w_t^M]^T$, which is a M-element vector with summation equal to $1$ at time $t$, we are able to achieve the re-weighted posterior probability sequence $\hat{P_t}$ as follows,
\begin{equation}
\label{equ:combination}
\hat{P_t}={\bf w}_t{\bf P}_t
\end{equation}
After the re-weighted combination, $\hat{P_t}$ is used for decoding.

\subsection{Attention Vector Estimation}

The attention vector can be estimated via evaluating the relative ASR performance between the microphone streams in unsupervised ways. Specifically, ASR performance measures are integrated to realize this purpose, stated as follows.

\subsubsection{Based on analysis of phoneme posteriorgram}

Researchers proposed to distinguish ASR performance through observing the relationship between recognition accuracy and representation of phoneme posteriorgram. Posterior distribution at a particular time point would converge to non-informative, as the signals were increasingly corrupted by noise or reverberation. Therefore, inverse entropy $1/H_i$ of $\overline{P}_t^i$ is a measure to determine the performance of microphone stream $i$ \cite{okawa1998multi}\cite{misra2002entropy}, so that the attention vector of each frame is given by 
\begin{equation}
\label{equ:entropy}
w_t^i=\frac{1/H_i}{\sum_{i=1}^M 1/H_i}
\end{equation}

By considering the temporal properties of phoneme posterior probability, mean time distance (M-measure) \cite{hermansky2013mean} and delta M-measure \cite{mallidi2015uncertainty} accumulate the divergences of probability estimates spaced over several time-spans. M-measure accumulates how similar or different every two probability vectors $P^i_{t-\Delta t}$ and $P^i_{t}$ are, by calculating their symmetric Kullback-Leibler divergence (KLD) $D(P^i_{t-\Delta t},P^i_t)$. If the speech were corrupted by stationary or slowly varying distortions, these distortions start dominating the signal and the phoneme posteriors become more similar, resulting in a lower average value of M-measure. Delta M-measure further takes phoneme dependence into account. Both M-measure and delta M-measure rely on long-term windows over hundreds of milliseconds. Stream with better ASR performance would have a larger value than the other streams in this window. Thus, a time-invariant attention vector having binary elements across the window is derived, which is given by $w_t^i==1$, if $M^i(\Delta t)>M^j(\Delta t)$, where $i\neq j$, $t$ belongs to all the frame time in the window.

\subsubsection{Based on unsupervised learning}

It's well-known that multi-layer neural network is good at modeling the complex data distributions. In the unsupervised learning approach, we use the autoencoder as an ASR performance monitor to model the output activations of DNN acoustic model \cite{mallidi2015autoencoder}.

In the training phase, an autoencoder is trained on the phoneme posterior sequences with Logit (to make the features more Gaussian) and principal component analysis (PCA) transformation (transformation basis of PCA is evaluated from the training data). The data for training the autoencoder is the same as that for training the DNN classifier. Mean square error (MSE) criterion is used.

In the test phase, the reconstruction error of test data is used as a measure of stream confidence, which means that a vector similar to the distribution of training data will yield a low reconstruction error compared to vectors drawn from a different distribution.
The lower the reconstruction error is, the better test and training data are matched, resulting in a better recognition accuracy. Different from M-measure and delta M-measure, autoencoder based ASR performance monitor is a frame-wise technique. The element of attention vector $w_n^i$ is given as follows,
\begin{equation}
\label{equ:autoencoder}
w_t^i=\frac{1/||e_i||^2}{\sum_{i=1}^M 1/||e_i||^2}
\end{equation}
where $||e_n||$ is the $l_2$ norm of reconstruction error vectors.

The temporal transition of phonemes is a speech-specific property, which is widely applied to the speech-related techniques. In this study, we use context-based phoneme posterior features centered by the current frame as the input, and current frame at time $t$ as the training target. To further relaxing the strict alignment of input features and corresponding targets and significantly reducing the input size \cite{peddinti2015time}, we exploit the TDNN structure with splices in the hidden layers to train the autoencoder.

%\subsection{Decode with the re-weighted posteriors}

\section{Experiment and Discussion}
\label{sec:experiment}

\subsection{Dataset and Baseline}

The proposed framework was evaluated on a subset of Mixer-6 dataset \cite{brandschain2010mixer}. During the recording, US English speakers were required to read from a list of sentences. 
In details, the recordings were conducted on-site at LDC in two distinct office rooms (denoted by "LDC" and "HRM" room) equipped with multi-channel recording platforms. Each room was set up with a matching set of 13 distinct microphones, placed at equivalent locations relative to the speaker. Therefore, this distributed setup meets our needs. 

The transcribed dataset was separated into training part and testing part for ASR experiments. For each utterance, we had synchronous (not time-aligned due to the propagation of the sound wave) 13 recordings simultaneously. We used the recordings from microphone 2 (head-mounted microphone, best acoustic channel) as the training data, and the remainders for testing.
Training data was 246.5 hours from more than 1350 speakers. And the test data consisted of two parts, one having 1031 utterances from 4 distinctive speakers in the "LDC" room and the other one having 898 utterances from another 4 speakers in the "HRM" room, respectively.

We tested all the 13 microphone streams on the typical DNN-HMM system trained on MFCC features, with 11 frames stacking (+5-5). To examine the improvement of the proposed scheme applied on the acoustic posteriors, the language model used for decoding was weak but equally for all the recognition experiments below. Table \ref{tab:baseline} shows the baseline WER for each microphone stream. %The WER varies among the streams under the influences of reverberation and environmental noises. 
Except for microphone 2, whose acoustic scene was matched with the training, we derived two test sets for the stream attention task. For the "LDC" set, we had twelve streams working in normal status. For the "HRM" set, ten streams worked well for ASR, however, the other two failed (Mic 3\&11). This phenomenon happens quite often in the real environments, as microphones might be out of charge suddenly or affected by strong noise and reverberation. The system should be robust in case of such microphone failures.

\begin{table}[htb]
  \caption{WERs(\%) of microphone streams (Mic X) on the two test sets. Recognizer was trained using the recordings from Mic 2 (Mic 2 was not used for testing).}
  \label{tab:baseline}
  \centering
  \begin{tabular}{c|cc}
    \hline
    \textbf{Stream Index} & \textbf{LDC room} & \textbf{HRM room} \\
    \hline
   	 Mic 1            & 23.8 & 27.0    \\
     Mic 2(Matched)   & {\bf 10.2} & 10.8    \\
     Mic 3            & 26.7 & 97.6    \\
     Mic 4            & 10.9 & {\bf 8.2}     \\
     Mic 5            & 12.9 & 12.9    \\
     Mic 6            & 10.1 & 8.7      \\
     Mic 7            & 15.1 & 15.3    \\
     Mic 8            & 14.0 & 12.6    \\
     Mic 9            & 22.7 & 18.3    \\
     Mic 10           & 11.3 & 13.4    \\
     Mic 11           & 10.6 & 75.9    \\
     Mic 12           & 14.6 & 12.7    \\
     Mic 13           & 19.9 & 21.9    \\
    \hline
  \end{tabular}
\end{table}

\begin{table}[tb]
  \caption{WERs(\%) comparison of various microphone stream re-weighting methods on the Mixer-6 multi-microphone dataset.}
  \label{tab:result}
  \centering
  \begin{tabular}{c|c|cc}
    \hline 
    {\textbf{Group}} & {\textbf{System\&Method}}  & \textbf{LDC} & \textbf{HRM} \\
    \hline
    \multirow{3}{*}{A} & Matched (Mic 2)                 & 10.1 & 10.8    \\
    				   & Best microphone (Oracle)        & 10.1 & 8.2     \\
                       & Lattice combination             & 11.7 & 19.3    \\                       
    \hline
    \multirow{2}{*}{B} & M-measure                       & 10.3 & 9.1     \\
                       & Delta M-measure                 & 10.2 & 8.8     \\           
    \hline
    \multirow{3}{*}{C} & Equal weights                   & 9.8  & 30.5    \\
                       & Inverse entropy re-weight       & {\bf 7.8}  & 7.9     \\
                       & AE re-weight w/o context        & 8.7  & 7.1     \\

    \hline
    \multirow{4}{*}{D} & AE re-weight w context [-8, 5]   & 8.5  & 7.1     \\
    				   & AE re-weight w context [-13,10] & 8.4  & 7.1     \\
                       & AE re-weight w context [-16,12] & 8.2  & {\bf 6.9}     \\
                       & AE re-weight w context [-20,14] & 8.6  & 6.9     \\
    \hline
    \multirow{2}{*}{E} & Inverse entropy Max             & 17.6 & 19.4     \\
                       & AE Max w context [-16,12]       & 20.8 & 18.2     \\ 
    \hline
  \end{tabular}
\end{table}

\subsection{Description of the comparative methods}
We compared the WER results between the proposed stream attention scheme using M-measure and delta M-measure \cite{hermansky2013mean}\cite{mallidi2015uncertainty} and the combination of lattices, generated by different streams by doing a union of the lattices \cite{fiscus1997post}. Both of them were processed sentence-by-sentence. 

We also took equal weights \cite{kittler1998combining}, inverse entropy \cite{misra2002entropy} and autoencoder (AE) \cite{mallidi2015autoencoder} for performance comparison since they were based on the frame-wise re-weighting of the HMM state posteriors in the proposed stream attention framework. What's more, as for the autoencoder hierarchy, we investigated the effect of using different temporal context sizes on WER. The autoencoder was trained with 6 layers (a 24-unit bottleneck layer in the middle), and each layer consisted of 512 Relu units. The temporal context was introduced via a TDNN architecture with different temporal resolution at each layer.

\subsection{Results}

Table.\ref{tab:result} shows the WER results using various comparative techniques. 
As shown in Group “A”, we pick out the matched case (Mic 2) and best microphone (oracle) as the baselines based on Table \ref{tab:baseline}. We can see that lattice combination performs worse than the baselines, especially on the "HRM" test set. Using the sentence-by-sentence strategy, our scheme carried out on the DNN posteriors show the superior performance to lattice level processing, which is delivered by Group "B". M-measure is able to make the system pay more attention to the best stream at the sentence level, but also can not outperform the best microphone stream. Delta M-measure slightly improves the selection accuracy. In some applications, the acoustic situation may change dynamically and solutions, which require such longer signal spans for making the stream selection, may not be appropriate. 

Group "C" gives the results of frame-wise re-weighting using different kinds of attention vectors. When applying equal weights to the twelve microphone streams, a better WER (9.77\%) is achieved on the "LDC" test set, which is superior to the best individual microphone. However, performance on "HRM" test set with two of the streams in bad condition gets much worse (30.45\%). In contrast, inverse entropy achieves a substantial improvement compared to the best microphone, showing a 22.8\% and 3.7\% relative improvement for "LDC" and "HRM" set in WER, respectively. But the WER improvement of "HRM" set is not as much as that of "LDC" set. This phenomenon does not occur when the autoencoder based attention vector was applied. We find that the improvements are consistent in both test sets. Furthermore, a trend can be observed by enlarging the context window, indicated by Group "D". The gain increases as we used more TDNN network context until [-16,12] (relative improvements for "LDC" set and "HRM" set are 18.8\% and 15.9\%, respectively) then the WER goes up when we apply a larger context [-20,14] on the "LDC" set. While results on the "HRM" set seem stable where only a little improvement has been achieved using more context information. 
According to the results, the entropy based system yields the best result on the "LDC" set, which does not include extremely corrupted microphone streams. The two extremely corrupted streams in the "HRM" set appear to be better dealt with using the autoencoder based system. 

\begin{figure}[tb]
  \centering
  \centerline{\includegraphics[width=8.5cm]{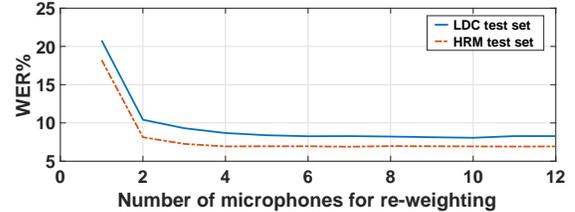}}
 \caption{WERs(\%) with respect to the number of microphone streams for the re-weighting scheme. Attention vector is calculated using the "AE re-weight with context [-16,12]".}
\label{fig:trend}
\end{figure}

To further gain insight into choosing the number of microphone streams in the frame-level re-weighting task, we explore the trend of WER via re-weighting the {\bf n-best} ($n=1,...,12$) streams. Group "E" shows an extreme case that only one microphone stream is locked given a frame (The "Max" means "Winner Takes All" for the total 12 streams). The results show a severe performance degradation for both the methods. However, if we focus on the trend in Fig.\ref{fig:trend}, we can find that the WERs decrease dramatically using only several microphone streams, and converge to steady with more streams. One interesting observation is that the "HRM" test set converges faster than the "LDC" set, which is in accord with the fact that fewer microphone streams have excellent WER results in the "HRM" set.

\section{Conclusion}
\label{sec:conclusion}

In this study, we aimed at improving the multi-channel far-field ASR performance by stressing the collaboration of microphone streams. A stream attention architecture was designed to give a more reasonable frame-wise fusion of HMM state posterior probabilities for the recognizer, regardless of the time misalignment of microphones. According to the ASR results on the Mixer-6 dataset, we found that our proposed framework showed a substantial capability to improve the performance with multiple inputs. The approach is highly parallel and, especially in the case of the entropy-based system, relatively computationally affordable. While the autoencoder system showed a more robust performance in case of microphone perturbation.

In future works, we would like to test the framework in the situation that the target speaker moves around and figure out the traces of active microphone streams. We are also interested in merging the posteriors using nonlinear networks.

%\section{Acknowledgement}
%The authors would like to thank Shinji Watanabe and Harish Mallidi at JHU for invaluable suggestions and discussion.

\vfill\pagebreak
\bibliographystyle{IEEEbib}
\bibliography{refs}

\begin{thebibliography}{10}

\bibitem{zhang2008maximum}
Cha Zhang, Dinei Flor{\^e}ncio, Demba~E Ba, and Zhengyou Zhang,
\newblock ``Maximum likelihood sound source localization and beamforming for
  directional microphone arrays in distributed meetings,''
\newblock {\em IEEE Transactions on Multimedia}, vol. 10, no. 3, pp. 538--548,
  2008.

\bibitem{markovich2015optimal}
Shmulik Markovich-Golan, Alexander Bertrand, Marc Moonen, and Sharon Gannot,
\newblock ``Optimal distributed minimum-variance beamforming approaches for
  speech enhancement in wireless acoustic sensor networks,''
\newblock {\em Signal Processing}, vol. 107, pp. 4--20, 2015.

\bibitem{meng2017deep}
Zhong Meng, Shinji Watanabe, John~R Hershey, and Hakan Erdogan,
\newblock ``Deep long short-term memory adaptive beamforming networks for
  multichannel robust speech recognition,''
\newblock in {\em Acoustics, Speech and Signal Processing (ICASSP), 2017 IEEE
  International Conference on}. IEEE, 2017, pp. 271--275.

\bibitem{fiscus1997post}
Jonathan~G Fiscus,
\newblock ``A post-processing system to yield reduced word error rates:
  Recognizer output voting error reduction (rover),''
\newblock in {\em Automatic Speech Recognition and Understanding, 1997.
  Proceedings., 1997 IEEE Workshop on}. IEEE, 1997, pp. 347--354.

\bibitem{xu2010improved}
Haihua Xu, Daniel Povey, Lidia Mangu, and Jie Zhu,
\newblock ``An improved consensus-like method for minimum bayes risk decoding
  and lattice combination,''
\newblock in {\em Acoustics Speech and Signal Processing (ICASSP), 2010 IEEE
  International Conference on}. IEEE, 2010, pp. 4938--4941.

\bibitem{waibel1989phoneme}
Alex Waibel, Toshiyuki Hanazawa, Geoffrey Hinton, Kiyohiro Shikano, and Kevin~J
  Lang,
\newblock ``Phoneme recognition using time-delay neural networks,''
\newblock {\em IEEE transactions on acoustics, speech, and signal processing},
  vol. 37, no. 3, pp. 328--339, 1989.

\bibitem{bahdanau2014neural}
Dzmitry Bahdanau, Kyunghyun Cho, and Yoshua Bengio,
\newblock ``Neural machine translation by jointly learning to align and
  translate,''
\newblock {\em arXiv preprint arXiv:1409.0473}, 2014.

\bibitem{kim2016recurrent}
Suyoun Kim and Ian Lane,
\newblock ``Recurrent models for auditory attention in multi-microphone
  distance speech recognition,''
\newblock in {\em INTERSPEECH}, 2016, pp. 3838--3842.

\bibitem{okawa1998multi}
Shigeki Okawa, Enrico Bocchieri, and Alexandros Potamianos,
\newblock ``Multi-band speech recognition in noisy environments,''
\newblock in {\em Acoustics, Speech and Signal Processing, 1998. Proceedings of
  the 1998 IEEE International Conference on}. IEEE, 1998, vol.~2, pp. 641--644.

\bibitem{misra2002entropy}
Hemant Misra, Herv{\'e} Bourlard, and Vivek Tyagi,
\newblock ``Entropy-based multi-stream combination,''
\newblock Tech. {R}ep., IDIAP, 2002.

\bibitem{hermansky2013mean}
Hynek Hermansky, Ehsan Variani, and Vijayaditya Peddinti,
\newblock ``Mean temporal distance: Predicting asr error from temporal
  properties of speech signal,''
\newblock in {\em Acoustics, Speech and Signal Processing (ICASSP), 2013 IEEE
  International Conference on}. IEEE, 2013, pp. 7423--7426.

\bibitem{meyer2016performance}
Bernd~T Meyer, Sri~Harish Mallidi, Angel Mario~Castro Mart{\'\i}nez, Guillermo
  Pay{\'a}-Vay{\'a}, Hendrik Kayser, and Hynek Hermansky,
\newblock ``Performance monitoring for automatic speech recognition in noisy
  multi-channel environments,''
\newblock in {\em Spoken Language Technology Workshop (SLT), 2016 IEEE}. IEEE,
  2016, pp. 50--56.

\bibitem{mallidi2015uncertainty}
Sri~Harish Mallidi, Tetsuji Ogawa, and Hynek Hermansky,
\newblock ``Uncertainty estimation of dnn classifiers,''
\newblock in {\em Automatic Speech Recognition and Understanding (ASRU), 2015
  IEEE Workshop on}. IEEE, 2015, pp. 283--288.

\bibitem{mallidi2015autoencoder}
Sri Harish~Reddy Mallidi, Tetsuji Ogawa, Karel Vesel{\`y}, Phani~S Nidadavolu,
  and Hynek Hermansky,
\newblock ``Autoencoder based multi-stream combination for noise robust speech
  recognition.,''
\newblock in {\em INTERSPEECH}, 2015, pp. 3551--3555.

\bibitem{peddinti2015time}
Vijayaditya Peddinti, Daniel Povey, and Sanjeev Khudanpur,
\newblock ``A time delay neural network architecture for efficient modeling of
  long temporal contexts,''
\newblock in {\em Sixteenth Annual Conference of the International Speech
  Communication Association}, 2015.

\bibitem{brandschain2010mixer}
L~Brandschain, D~Graff, C~Cieri, K~Walker, C~Caruso, and A~Neely,
\newblock ``The mixer 6 corpus: Resources for crosschannel and text independent
  speaker recognition,''
\newblock in {\em Proc. of LREC}, 2010.

\bibitem{kittler1998combining}
Josef Kittler, Mohamad Hatef, Robert~PW Duin, and Jiri Matas,
\newblock ``On combining classifiers,''
\newblock {\em IEEE transactions on pattern analysis and machine intelligence},
  vol. 20, no. 3, pp. 226--239, 1998.

\end{thebibliography}

\end{document}